\documentclass[10pt,twocolumn]{article}

\usepackage[utf8]{inputenc}
\usepackage[T1]{fontenc}

\usepackage{graphicx}

\usepackage{lipsum}
\usepackage{xcolor}

\usepackage{authblk}
\usepackage[margin=2cm]{geometry}
\usepackage{mathtools, cuted}

\usepackage{url}
\usepackage{bm}
\usepackage{amsfonts}
\usepackage{amsmath}
\usepackage{amssymb}
\usepackage{mathrsfs} 
\usepackage{siunitx}

\title{Role of evaporation in stability of foam films and foams}
\author[1]{François Boulogne}
\author[1]{Emmanuelle Rio}
\affil[1]{Université Paris-Saclay, CNRS, Laboratoire de Physique des Solides, 91405, Orsay, France.}

\date{\today}
\begin{document}

\twocolumn[
    \begin{@twocolumnfalse}
        \maketitle
        \begin{abstract}
            Understanding the stability of foam films and foams remains a challenge, despite the considerable efforts provided by the scientific community to refine their physical descriptions.
            This persistent difficulty underscores the interplay of various complex factors.
            Recently, the role of evaporation has attracted attention for its dual impact: it can either enhance stability or accelerate bursting.
            To depict a comprehensive overview on soapy objects, we propose first a short description on the evaporation of drops to present some key results on the evaporation-induced cooling and the consequences on the internal flows generated by Marangoni effects.
            Then, we review the literature on foam films and foams, examining three distinct systems: pure liquids, non-volatile, and volatile surface active molecules.
            We show that evaporation of foam films and foams can lead to significant variations of temperature.
            In conclusion, we identify a series of open questions and suggest potential research pathways to address these challenges.
        \end{abstract}
    \end{@twocolumnfalse}
]

%
%
\section{Introduction}

Foams are a broad category of materials, and they are commonly found throughout our surroundings. 
For example, solid foams can be found in construction materials, food products, and cosmetics, while liquid foams are used as cleaning products and fire extinguishing agents. 
The low amount of material required, their light weight, as well as their exceptional mechanical, thermal, and acoustic properties give foam a distinct advantage for applications requiring low cost, high efficiency, and low environmental impact. 
Solid foams are produced by solidifying a liquid phase. 
Liquid foams are generally unstable, as their equilibrium is associated with two separate phases: liquid and gas. Therefore, learning how to
control the stability of foams and ultimately that of foam films is one of the key questions driving both applied and fundamental research \cite{Bikerman1973,Stevenson2012,Pugh2016}.

Different mechanisms are identified in the literature as contributors to foam aging.
First, gravity plays a significant role in the foam aging process by causing gravitational drainage, which leads to a separation of the liquid and gas phases in a foam due to their density difference.
Capillary effects counteract this drainage to some extent, preventing complete liquid loss.
This process results in an equilibrium with a wetter foam at the bottom and drier foam at the top, which creates spatial variations of the mechanical properties of the material \cite{Wang2016}.

A second phenomenon participating directly in foam aging is called coalescence.
A coalescence event is the rupture of a film separating two bubbles leading to a single bubble.
As this process can occur in avalanches, the number and size of bubbles constituting the foam may vary significantly.
If some event occurs at the foam interface, the volume of foam will be reduced.
Understanding the causes of the film rupture remains a challenge that combines effects of the capillary pressure, surface area of foam films, thermal fluctuations, and also evaporation \cite{Langevin2019}.

A third process, named coarsening or Ostwald ripening is associated with  the permeability of foam films to the gas contained in bubbles \cite{SaintJalmes2006}.
Small bubbles having a larger Laplace pressure than bigger ones, the smaller bubbles empty and disappear, reducing the total number of bubbles and increasing the bubble size.

The different phenomena that we just mentioned are nearly exclusively studied in an atmosphere saturated with the vapor of the involved liquid by the means of an enclosing case.
In other words, evaporation is prevented to avoid an additional transport mechanism.
However, in practice, evaporation is relevant in many applications and undoubtedly plays a role on the stability of foam films \cite{Tobin2011,Pasquet2022a} and foams \cite{Li2012a}.

Some studies are devoted to describing the role of evaporation on foam films and foam stability, which is the topic of the present article.
As we will see, the number of studies is limited and a comprehensive picture has not been established yet.
Indeed, the interest of the community in these questions started to grow recently and we think this is an appropriate time to depict the state of the art and identify some challenges.
More precisely, this article is motivated by recent measurements on the temperature of evaporating foam films and foams, which can be crucial to decipher the enigma of the foam film lifetime.

Insightful results are provided by studies on the evaporation of drops, for which the thermodynamics, the hydrodynamics, and the role of the physical-chemistry are explored.
These results will serve to present the main concepts.
Thus, in Section~\ref{sec:drops}, we introduce some key results on the Marangoni flows in evaporating droplets, without pretending to perform a comprehensive review.
Next, in Section~\ref{sec:thinning}, we present the main findings on the foam film and foam stability where evaporation is at play.
This Section is organized by systems to highlight their specificity and we will demonstrate that although the evaporative cooling is sometimes mentioned, the temperature has not been directly measured.
Then, in Section~\ref{sec:temperature},
we start with a presentation of psychrometry to highlight the significance of heat radiative transfer in the prediction of the temperature.
We use these concepts to summarize recent results we obtained  where we focused on the  temperature of evaporating foam films and foams.
Finally, we propose a series of key questions and promising directions for future work in Section~\ref{sec:perspectives}.

%
%

\section{Marangoni flows in evaporating drops}\label{sec:drops}

Although the renewed interest regarding the consequences of evaporation on foam films leading to thermal and solutal Marangoni flows is relatively recent, a strong background has been established by the community working on droplet evaporation.
Thus, we present some key results from this community that are useful for understanding the evaporation phenomena that occur in foam films and foams.

\subsection{Thermal Marangoni flows}

\begin{figure}[ht]
    \centering
    \includegraphics[width=.9\linewidth]{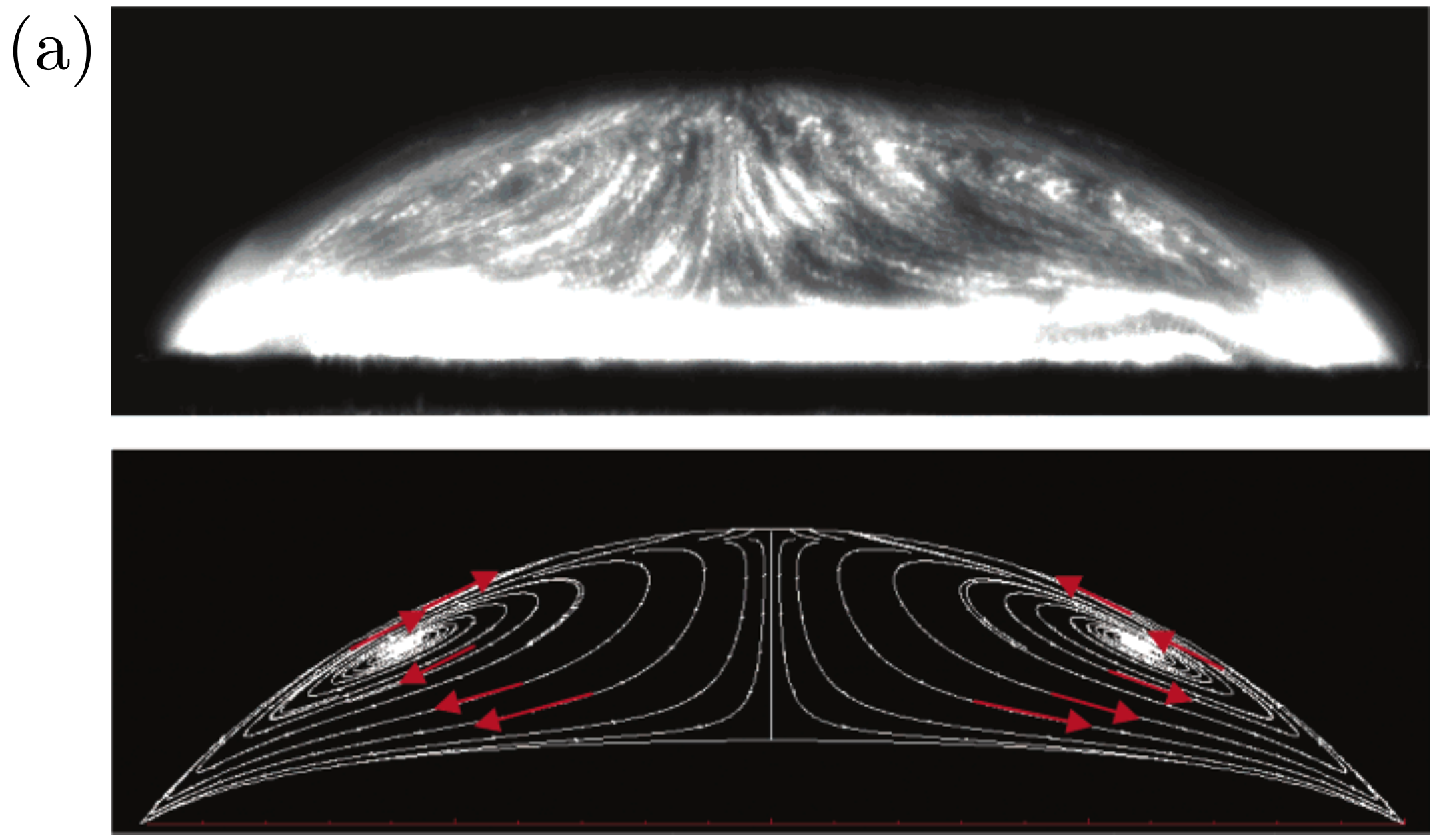}
    \includegraphics[width=.9\linewidth]{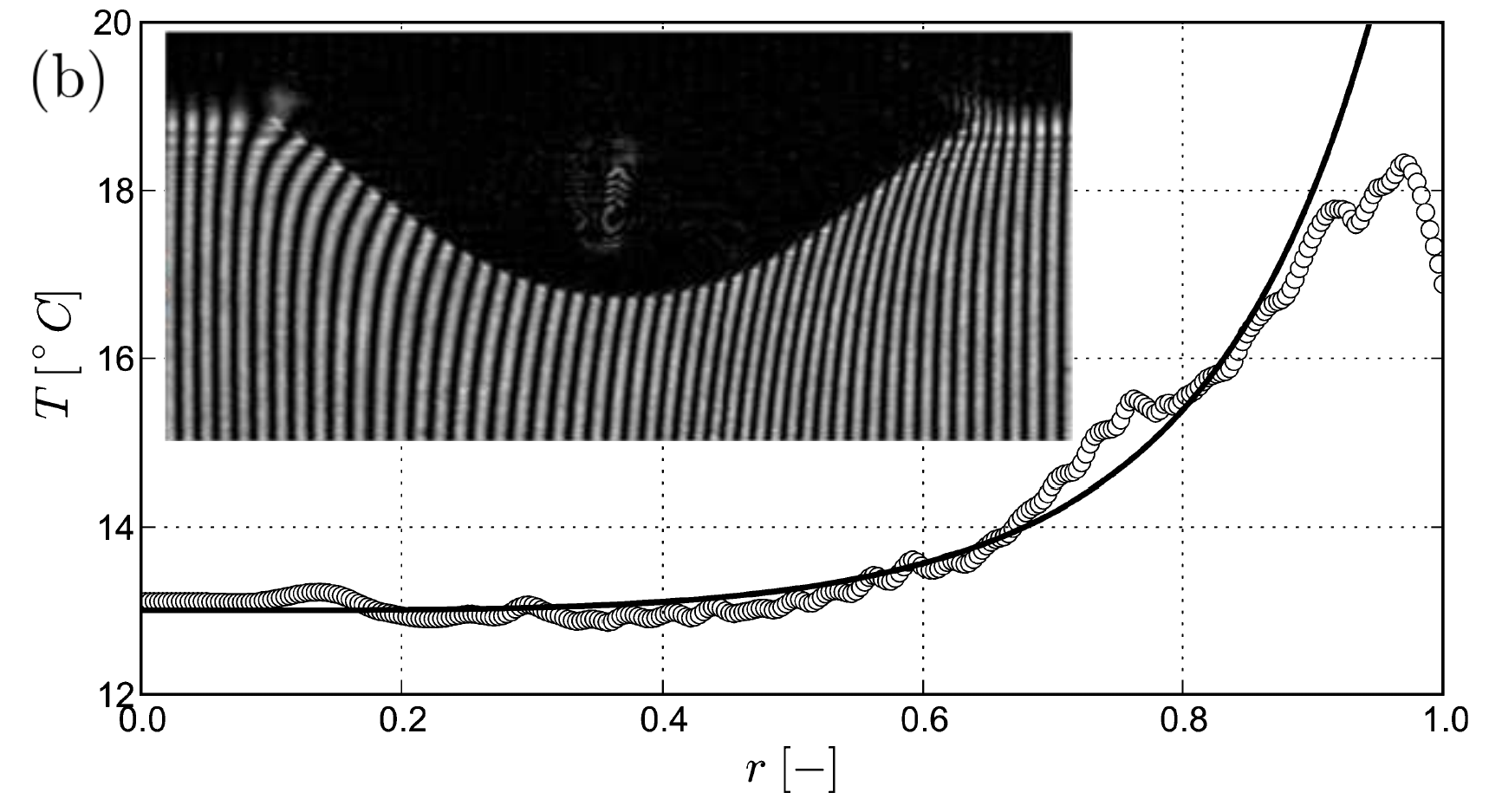}
    \caption{(a) Side views of an evaporating sessile drop of octane.
The top image corresponds to the experimental observation of the motion of tracers while the bottom picture is the predicted flow field.
The vortex is caused by the thermal Marangoni stress at the interface.
These images are taken from Ref.~\cite{Hu2006}.
(b) Temperature of the interface of an evaporating pendant drop of a commercial liquid, deduced by interferometric measurements performed at an ambient temperature of $24~^\circ$C.
The image illustrates the drop with the interferometric fringe pattern.
The coordinate $r=0$ corresponds to the apex of the drop and $r=1$ to the contact line.
This measurement highlights both the cooling effect and the temperature gradients.
This graph has been adapted from Ref.~\cite{Dehaeck2014}.
}
    \label{fig:drops}
\end{figure}

A spherical droplet is considered as the simplest geometry for modeling evaporation.
Langmuir proposed in 1918  a calculation based on the assumption of a transport process limited by the diffusion of vapor in the atmosphere to explain that the evaporation rate is proportional to the radius of the sphere \cite{Langmuir1918}.
The cooling effect was included in this calculation by Fuchs in 1947 where both the thermal conduction and the radiation are considered in a quiescent atmosphere \cite{Fuchs1947}.
Due to the variation of the vapor pressure with the temperature, this effect implies a lower evaporation rate.
Next, imposing an air flow is an additional complexity addressed by Ranz and Marshall who considered the evaporation of water drops hanging at the tip of a capillary tube in a convective air flow \cite{Ranz1952,Ranz1952a}.
In this study, they used the equations proposed by Frossling \cite{Frossling1938} to model the effect of the air velocity on the transfer rate of heat and mass.
These foundational works have served for instance in atmospheric science to understand the evaporation of sea spray \cite{Andreas1989,Andreas1995}, requiring numerical tools to obtain the temperature and the lifetime of drops \cite{Sobac2015}.
More recently, evaporation of small droplets regained interest in the context of air-transmitted viruses \cite{Netz2020,Netz2020a}, where analytical solutions have been derived with assumptions on the variation of the saturated vapor pressure with the temperature \cite{Corpart2023a}.
Thus, the fundamentals of the evaporation of spherical drops appears to be well documented.
Another particularly well-studied geometry is a drop partially wetting a surface, a sessile drop.

After the famous paper by Deegan \textit{et al.} in 1997 on the stain left by an evaporating sessile drop containing particles \cite{Deegan1997}, the interest of the community for the evaporation of drops has grown considerably.
In most of the studies devoted on the coffee-stain effect, the evaporation of droplets is considered as an adiabatic process.
Particles are transported toward the contact line due to a radial flow of the liquid imposed by the evaporation, but for which the diverging nature of the evaporating flux at the contact line is not essential \cite{Boulogne2017a,Dambrosio2023}.
Nevertheless, evaporation-induced cooling and temperature gradients can have significant effects both on the droplet lifetime and the motion of the particles.
Deegan noticed that temperature gradients in sessile drops can trigger Marangoni flows due to spatial variations of the surface tension \cite{Deegan2000}.
A direct visualization of vortices in the liquid phase of an evaporating drop has been performed by Hu and Larson as shown in figure~\ref{fig:drops}(a) \cite{Hu2006} .
Additionally, the temperature of the interface of a sessile drop has been obtained from an interferometric measurement of the refractive index field in the vapor phase above the interface, which is presented in figure~\ref{fig:drops}(b)  \cite{Dehaeck2014}.
The theoretical challenge to predict the temperature field in sessile drops is to properly describe the role of the substrate in the heat exchange.
Ristenpart \textit{et al.} demonstrated experimentally that the magnitude of Marangoni flows depends on the ratio of the thermal conductivities of the substrate and the liquid \cite{Ristenpart2007}.
The associated model predicts that the value of this ratio modifies the orientation of the temperature gradient, although this has not been observed yet \cite{Gelderblom2022}.
Indeed, for pure liquids, the predictions on the induced internal flows strength are greater than the experimental observations, which tends to indicate that contaminant must play a significant role through solutal Marangoni effects competing with thermal induced flows \cite{Gelderblom2022}.

\begin{figure}
    \centering
    \includegraphics[width=\linewidth]{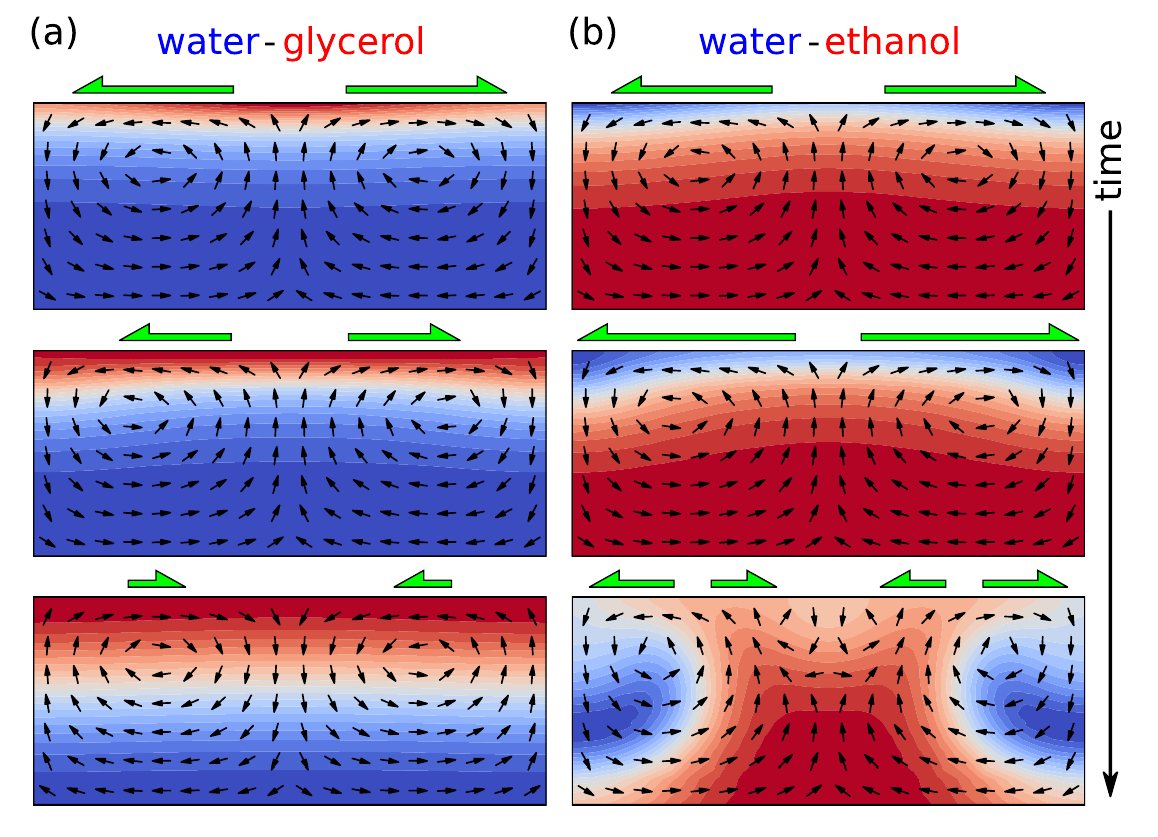}
    \caption{Illustration of solutal Marangoni flows caused by the evaporation of binary mixtures of (a) water-glycerol and (b) water-ethanol.
    Regions rich in water are colored in blue, and poor in water in red.
    Black arrows show the flow field.
    The evaporation creates spatial gradients of the liquid composition due to the difference of volatility.
The top to bottom sequence indicates the time evolution that shows either a self-inhibiting Marangoni flow for the glycerol solution or the emergence of a chaotic flow pattern in the case of a water-ethanol mixture.
The figure is extracted from Ref.~\cite{Gelderblom2022}.}
    \label{fig:marangoni}
\end{figure}

\subsection{Solutal Marangoni flows}

Evaporation of binary mixtures or more complex solutions are also prone to trigger Marangoni flows.
Here, the flow is caused by concentration gradients affecting the surface tension.
For instance, the presence of surface active molecules modifies significantly the internal liquid flow \cite{Marin2016} that has a strong impact on the transport of particles in an evaporating sessile droplet \cite{Still2012}.
Marangoni stresses can be sufficiently significant to affect the shape of the free interface as observed with mixtures of alkanes \cite{Guena2007a} but also the famous phenomenon called \textit{tears of wine} \cite{Fournier1992}.
Other alcoholic beverages such as ouzo \cite{Diddens2017} and whisky \cite{Kim2016} also inspired studies by the richness of flow patterns induced by the coupling of evaporation and Marangoni stresses.
A key role of the liquid composition is illustrated in figure~\ref{fig:marangoni} where a solutal Marangoni flow can either dampen itself or, in contrast, become chaotic.
Indeed,  the liquid composition at the interface can modify the local evaporation rate, the surface tension gradient, and the local liquid density.
Recently, Diddens \textit{et al.} solved theoretically the complex problem of the evaporation of a binary mixture by considering the aforementioned mechanisms for sessile and pendant drops (inverted orientation of gravity) \cite{Diddens2021}.
They showed that the interplay between these effects, neglecting thermal effects induced by evaporation, already leads to five different flow patterns.
Thus, the evaporation of complex systems where solutal Marangoni flows are at play can lead to a variety of observations strongly connected to the physical-chemistry of the system, further increasing the modeling complexity.



\section{Evaporation of films and foams}\label{sec:thinning}

\subsection{Pure liquid films}\label{ssec:pure_film}

Although this review is focused on foam films and foams, some results on pure liquid films are particularly enlightening, so that we will start by considering them.
Joseph Plateau reported observations and commented on the stability of surface bubbles with respect to a large variety of chemical compositions \cite{Plateau1873}.
Notably, in water, the bubble lifetime was found to be significantly increased by decreasing the relative humidity, which means that evaporation has a stabilizing effect.
This observation is, at first sight, counter-intuitive from the fact that film bursting is related to its thinning.

\begin{figure}[h!]
    \centering
    \includegraphics[width=.5\linewidth]{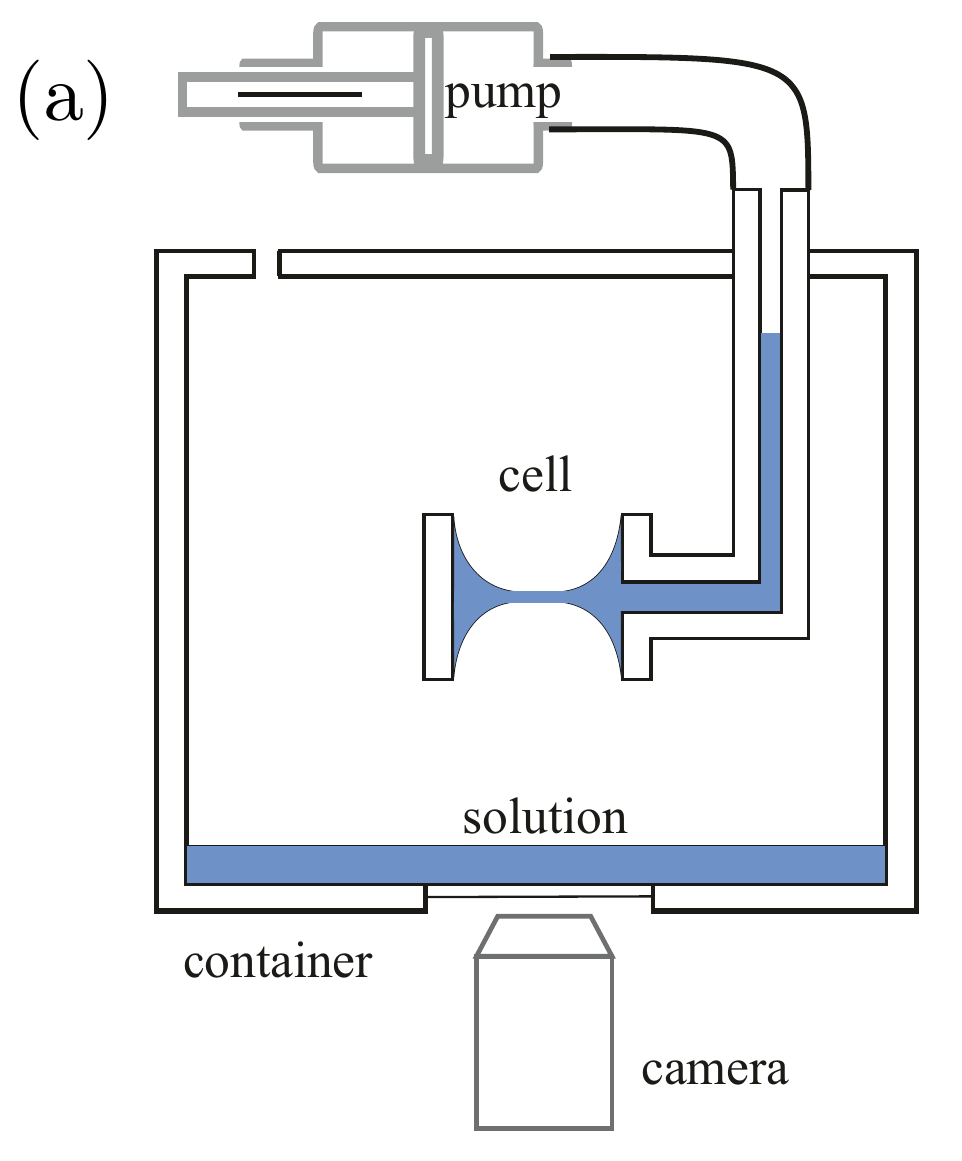}
    \includegraphics[width=.9\linewidth]{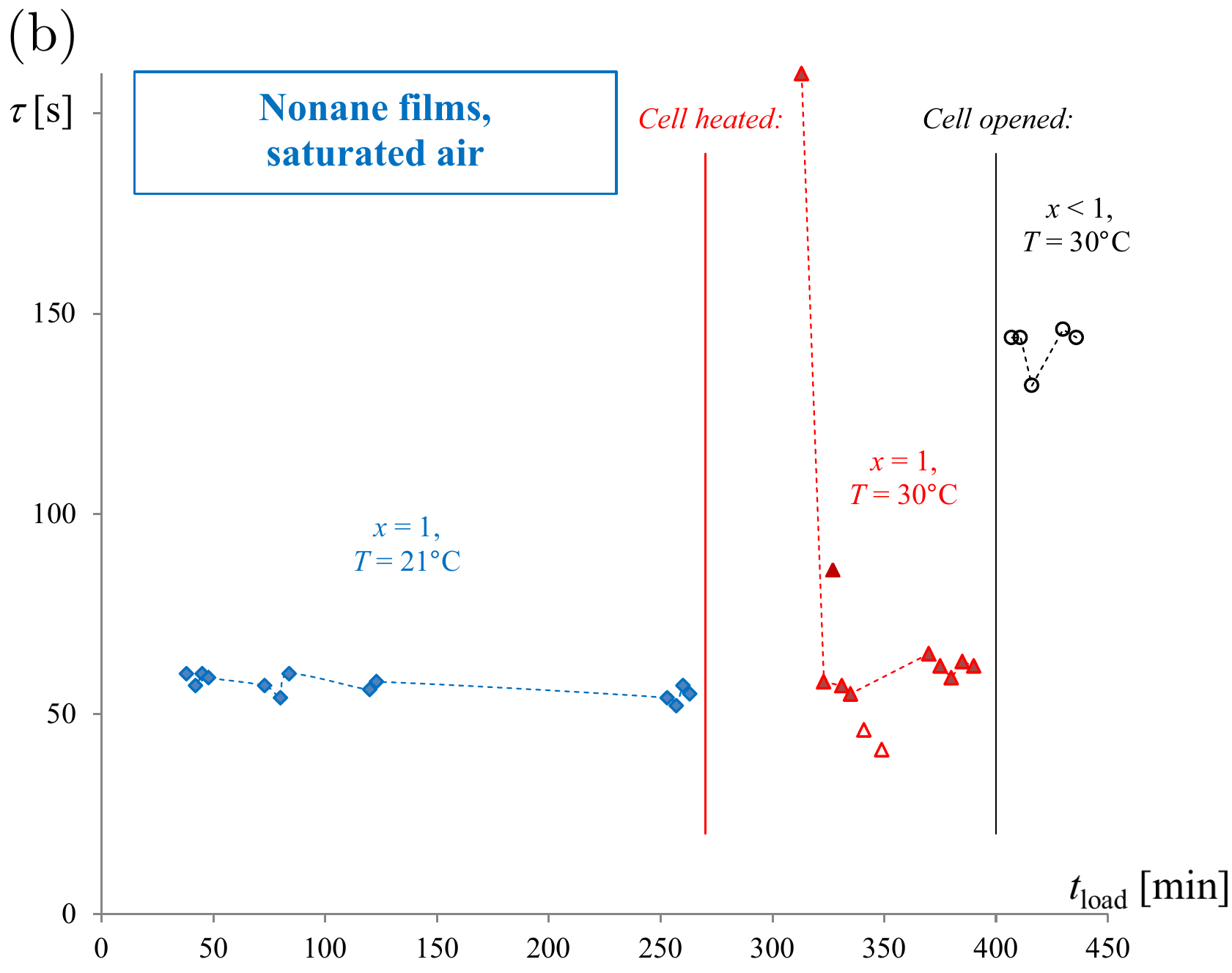}
    \caption{(a) Schematic of a thin film balance in a controlled atmosphere.
    The liquid at the bottom of the container is placed to create a saturated atmosphere.
    (b) Lifetime $\tau$ of films of nonane as a function of $t_{\rm load}$, the time after loading the cell.
Blue points are obtained in a saturated atmosphere ($x=1$).
The red vertical line corresponds to an increase of the temperature.
The first red point is obtained while equilibrium is not yet reached, such that evaporation proceeds.
When the atmosphere is back to saturation ($x=1$), the other red points recover values similar to the previous experiments in blue.
Black points are obtained with evaporation ($x<1$), showing an extended lifetime.
The schematic and the graph are adapted from Ref.~\cite{Novev2017}.}
    \label{fig:TFB}
\end{figure}

This stabilizing effect is corroborated by modern experiments on a thin film balance for instance.
This experimental setup, originally proposed by Sheludko \cite{Sheludko1967}, offers the possibility to measure or impose the liquid pressure in a thin film (Fig. \ref{fig:TFB}(a)).
Yaminsky \textit{et al.} reported that films of ethanol, pentane, and chloroform are more stable with evaporation than without \cite{Yaminsky2006}.
They also observed that a chaotic flow occurs in the film, which depends on the liquid volatility and the partial pressure of vapor in the atmosphere.
Although no experimental measurement of the temperature field has been performed in this system, this flow is attributed to the cooling induced by the evaporation.
In this geometry, the temperature in the center of the film is supposed to be lower than at the edge, which is near the environmental temperature due to the substrate.
The temperature gradients trigger a stabilizing Marangoni flow toward the center of the film.
In addition, similar observations have been made with pure water films, where the thinning rate can be imposed by a variation of liquid pressure \cite{Yaminsky2010}.
If the thinning rate is quasi-static, an effective stabilization is obtained, whereas a faster thinning rate leads to a premature rupture due to the hydrodynamics resistance in the film.

An experimental illustration of the stabilizing effect is provided in figure {\ref{fig:TFB}}(b).
After loading the container with nonane to saturate the atmosphere, they repetitively measured the lifetime of noname films at $21^\circ{\rm C}$.
Then, they increased the temperature to $30^\circ{\rm C}$, and after equilibration, the lifetime is similar to the one obtained at $21^\circ{\rm C}$.
Finally, they opened the container to reduce the vapor pressure of nonane and they observed that the lifetime is about three times longer.
A theoretical study associated with these thin film balance experiments has been proposed by Novev \textit{et al.} \cite{Novev2017}.
Assuming that the temperature follows a quadratic function of the radial position, they obtained the thinning rate from hydrodynamic equations from which several conclusions can be drawn.
The stability is enhanced as the evaporation rate is increased and for larger film radii (Fig. \ref{fig:TFB}(b)).
The model also predicts a critical thickness below which the Marangoni flow is not sufficient to stabilize the film.
These conclusions are qualitatively supported by additional experimental measurements on water and alkanes.

The role of impurities has been reported by Yaminksy \textit{et al.}  \cite{Yaminsky2010a}, showing that they play a significant role in the stability only if the film evaporates.
The stability of water film can be decreased when the contaminant is an organic compound, and conversely, the stability is enhanced by the addition of salt.
The difference lies in the surface tension that decreases with an increasing concentration of an organic compound, but increases with salt.
Thus, the impurities are introducing a solutal Marangoni flow, competing or reinforcing the thermal Marangoni flow.

These conclusions are also drawn with surface bubbles.
For instance, images obtained with an infrared camera support the existence of temperature gradients in surfactant-free surface bubbles \cite{Menesses2019}.
Additionally, the film thickening due to the transport of salt is also measured in this geometry \cite{Poulain2018}.
This thickening results in enhanced stability \cite{Miguet2021}. 

\subsection{Non-volatile surfactant}

\begin{figure}
    \centering
    \includegraphics[width=\linewidth]{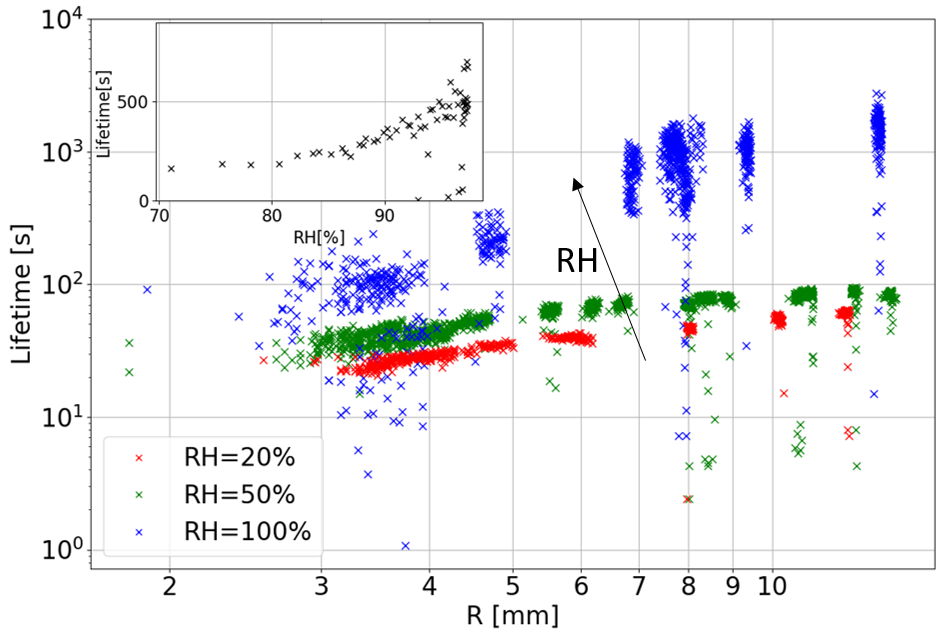}
    \caption{Lifetime of surface bubbles produced in a solution of surfactant (TTAB).
    Data are represented as a function of the bubble radius for three different relative humidity values in the main plot.
    The inset shows the lifetime as a function of the relative humidity.
    The figure is from Ref.~\cite{Miguet2019}.}
    \label{fig:surface}
\end{figure}

The number of studies devoted to the role of evaporation on foam films and foams is surprisingly limited.
First, we focus our attention on foam films stabilized by soluble and non-volatile surfactant molecules.
Evaporation contributes to the film thinning by transfer of solvent molecules from the film to the atmosphere.
Champougny \textit{et al.} generated vertical foam films by pulling a frame out of a surfactant solution and they measured the length at which the film ruptured by varying systematically the relative humidity of the atmosphere.
Decreasing the relative humidity leads to more fragile films \cite{Champougny2018}.
The dynamics of the film thickness can be rationalized by combining the gravitational drainage and evaporation.
As the thickness decreases, the viscous drainage slows until evaporation becomes significant in the thinning process, ultimately causing the rupture.

Similar conclusions have been made on the lifetime of bubbles floating at the surface of a surfactant solution  where the drainage is ensured either by capillarity or by gravity depending on the bubble size \cite{Poulain2018,Miguet2019,Miguet2021}, as shown in figure \ref{fig:surface}.
In this situation, the lifetime can be rationalized taking into account drainage and evaporation.
Nevertheless, Pasquet \textit{et al.} made additional observations by performing measurement of the thinning rate of surface bubbles of TTAB (Tetradecyltrimethylammonium bromide) \cite{Pasquet2022}.
Without evaporation, they recorded no effect of the TTAB concentration either below or above cmc.
But, by letting the bubble evaporate with solutions of concentration larger than cmc, they observed a delay in the film thinning.
These observations have not been rationalized so far.

To limit evaporation, a common practice used by artists and also parents with their children, is to add glycerol  to the surfactant solution \cite{Pasquet2022a}, which is a non-volatile and hygroscopic compound, more viscous than water.
As evaporation proceeds, the concentration of glycerol increases in the foam film until an equilibrium is reached between the water concentration in the film and the vapor concentration in the atmosphere.
Roux \textit{et al.} demonstrated that armored bubbles made of small beads bound with a water-glycerol mixture can be maintained forever because the liquid is retained by capillarity in the granular media and evaporation is prevented \cite{Roux2022}.
This technique has been proved to be efficient on surface bubbles stabilized with surfactants where the consideration of the chemical potential of the solution was sufficient to describe the film thinning \cite{Pasquet2022}.

Besides foam films and bubbles, some experimental observations indicate also that foams are less stable due to evaporation.
Li \textit{et al.} considered two situations: the time evolution of a static foam height \cite{Li2012a} and the maximum height reached by evaporating foams produced from a continuous injection of gas \cite{Li2010}.
They interpreted that evaporation creates a solutal Marangoni flow that tends to drain the liquid from the free surface of the foam, which adds to the thinning caused by evaporation \cite{Li2012a}.

\subsection{Volatile surface active molecules}

The case of volatile surface active molecules on the film stability has been discussed earlier in the literature.
Talmud and Suchowolskaya reported in 1931 experimental observations on the lifetime of such films on vertical frames, indicating a stabilizing effect of evaporation \cite{Talmud1931}.
This result is opposed to the behavior of non-volatile surfactant molecules discussed in the previous paragraph, but has been confirmed  by several authors with different systems \cite{Jones1967,Pigeonneau2012,Lorenceau2020}.
Neville and Hazlehurst qualitatively interpreted these results by suggesting a role of the cooling induced by evaporation coupled with the thickness variation due to the gravitational drainage.
For them, the thinner part of the film is expected to be at the lowest temperature because this region has a lower heat capacity due to its lower thickness and also a lower thermal conductivity limiting the heat flux from the rest of the film.
Thus, they concluded that a thermal Marangoni flow feeds the upper part of the film limiting the drainage, and stabilizing the film.

Additional experiments where mixtures of non-volatile surfactant with alcohol were performed in 1967 by Jones and Mysels on vertical films, for which they proposed a different interpretation \cite{Jones1967}.
Evaporation of alcohol leaves a surface of higher surface tension, which generates a solutal Marangoni flow in the same manner as the tears of wine \cite{Fournier1992}.
This interpretation is also adopted by Pigeonneau \textit{et al} on a different system that consists of a molten glass where some surface active molecules also evaporate \cite{Pigeonneau2012}.

Interestingly, the role of temperature gradients has been mentioned as a possible effect in earlier works, as long as a solutal Marangoni.
Experiments where a temperature gradient is imposed on a surface bubble of a pure solution also demonstrate a stabilizing effect \cite{Poulain2018,Nath2022}.

\subsection{Partial conclusion}
To summarize,  evaporation can have, for pure liquid films, a stabilizing effect by creating a temperature gradient responsible for a Marangoni flow compensating for the liquid loss by evaporation.
This mechanism can be altered by the presence of contaminants, either reduced or reinforced depending on their nature.
For films with non-volatile surfactants, evaporation is found to reduce the stability due to the film thinning.
In contrast, for volatile surface active molecules, evaporation produces a stabilizing effect.
However, the discrimination between the solutal and the thermal Marangoni flow has not been performed yet.

%
%
\section{The temperature of films and foams}\label{sec:temperature}

Some results from the literature exposed in Section~\ref{sec:thinning} suggest that the temperature of evaporating films is one of the key attributes in the stability of these systems, although an in-depth consideration has not been performed on the stability of soapy objects.
Two main aspects can be identified:
\begin{itemize}
    \item The evaporation rate depends on the temperature.
    This dependency mainly comes from the variation of the vapor pressure, at least for water \cite{Corpart2023b}.
    Therefore, a quantitative description of the evaporation rate must include the role of the enthalpy of vaporization.
    \item Temperature gradients lead to surface tension gradients, which drive liquid flows.
    These liquid flows modify the liquid distribution through variations of the liquid film thickness, which is linked to the stability.
    In addition, these flows modify also the liquid composition in the case of mixtures, further complicating the description.
\end{itemize}
Therefore, we estimate that understanding the temperature of soapy objects is a crucial interest to progress on the description of the foam film rupture.
The approach we propose is to introduce psychrometry that will provide a framework to predict the temperature of foam films and foams.

\subsection{Psychrometry}\label{sec:psychro}

\begin{figure}
    \centering
    \includegraphics[width=.45\linewidth]{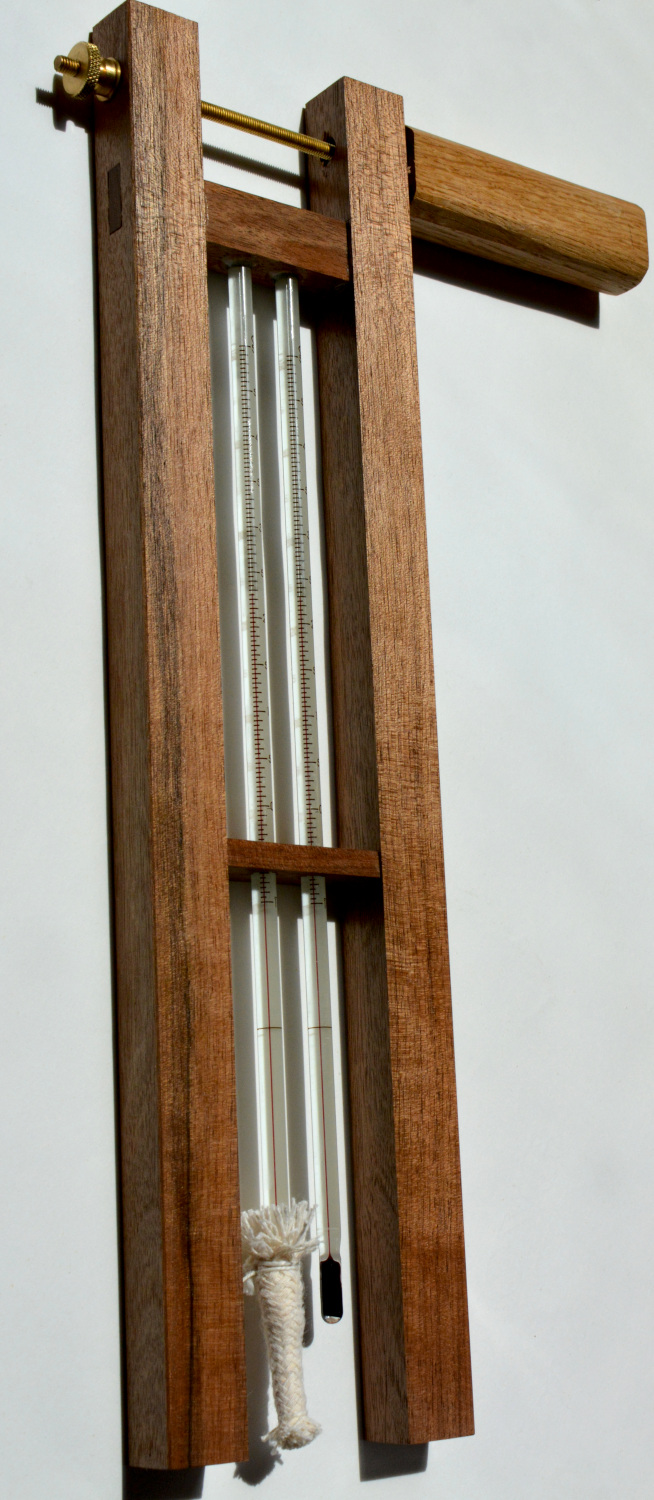}
    \includegraphics[width=.45\linewidth]{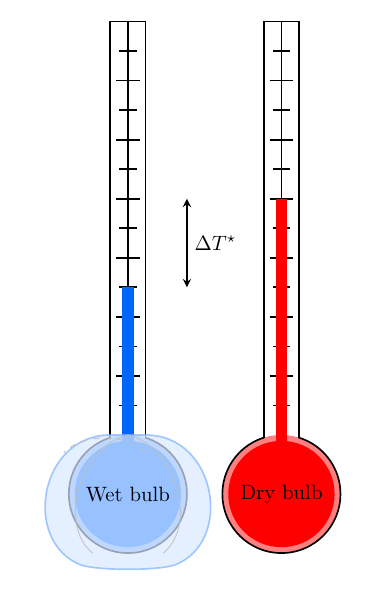}
    \caption{Photograph and schematic of a psychrometer.
Two thermometers are placed side by side, the one on the right with a dry bulb, and the one on the left with a mesh to wet the bulb.
The temperature difference $\Delta T^\star$ is measured.
    }
    \label{fig:psychrometer}
\end{figure}

Psychrometry, whose etymology comes from Greek \textit{psuchron} meaning ``cold'', and \textit{metron} meaning ``means of measurement'', is a branch of thermodynamics focused on the liquid-vapor mixtures.
The related instrument, invented by the end of the 18th century \cite{Playfair1997,Ivory1822}, the psychrometer, meant to measure the atmospheric humidity, is composed of two thermometers where one bulb is kept dry while the other one is maintained wet (Fig.~\ref{fig:psychrometer}).
The wetness of the wet bulb is ensured by a mesh or a muslin pre-moistened with water.
Due to evaporation, the wet bulb thermometer indicates a lower temperature than the dry bulb temperature.
Psychrometer is a type of hygrometer, from which the temperature difference between the bulbs is converted to a relative humidity.

Evaporation of the wet bulb induces a temperature decrease due to the enthalpy of vaporization $\Delta_\text{vap}H$.
Therefore, the temperature difference can be related to the evaporation rate, which depends on the relative humidity, which puts the psychrometer in the category of hygrometers.
Understanding the physics underlying this instrument can serve as a foundation for modeling evaporating liquids.

Let us begin with a description of the physics principle \cite{Corpart2023b}.
A psychrometer is operated in a steady state regime, which means that the energy required for vaporizing the liquid is balanced by the heat flux.
The heat flux can be transmitted by thermal diffusion and radiation.
Experimental observations indicate that blowing air on the psychrometer decreases the temperature difference up to a certain velocity.
The reason for this characteristic velocity comes from the difference between transfer by diffusion and radiation.
Both the evaporation and the thermal diffusion increase with the air velocity due to the apparition of a boundary layer across which the transfers are done.
However, the air flow has no effect on the radiative process.
Thus, the air flow tends to decrease the relative contribution of radiation in the energy balance up to the point that it can be neglected.
In addition, both the thermal conduction and the evaporative flux have the same variation with the air velocity, which are canceling out above the characteristic air speed.
Consequently, to get the most reliable measurements, psychrometers are ventilated such that the temperature difference becomes independent of the air velocity.

The relative humidity at a temperature $T$ is defined as the ratio of the vapor pressure $p$ and the saturated vapor pressure $p_{\rm sat}$:
\begin{equation}\label{eq:def_RH}
    \mathcal{R}_{H}(T)=\frac{p(T)}{p_\text{sat}(T)}.
\end{equation}
The psychrometer relates a temperature difference $\Delta T^\star=T_\text{dry}-T_\text{wet}$ measured between the dry and the wet bulb respectively, and a pressure difference $\Delta p^\star = p(T_{\rm dry}) - p_{\rm sat}(T_\text{wet})$.
The so-called psychrometric equation is a phenomenological equation stating that the pressure difference is proportional to the temperature difference $\Delta T^\star$, the atmospheric pressure $P$, and a psychrometer coefficient ${\cal A}$, which reads:

\begin{equation}\label{eq:psychrometric_equation}
    \Delta p^\star  =   - {\cal A} P \Delta T^\star.
\end{equation}

Modeling the wet bulb as a sphere of radius $R$ for the sake of simplicity, we can demonstrate that the coefficient ${\cal A}$ can be approximated as

\begin{equation}\label{eq:psychrometer_coeff_model_approx}
    {\cal A} \simeq  \frac{\lambda_{\rm air} M_{\rm air}}{ \Delta_\text{vap}H \, {\cal D}_{\rm w} \rho_{\rm air} }  \left( 1 + \frac{ T_\text{dry}^3 }{T_{\rm c}^3} \right),
\end{equation}
where $\lambda_{\rm air}$, $M_{\rm air}$, and $\rho_{\rm air}$ are the thermal conductivity, the molar mass and the density of air, respectively.
The diffusion coefficient of vapor in air is denoted ${\cal D}_{\rm w}$.
In equation \ref{eq:psychrometer_coeff_model_approx}, the critical temperature $T_{\rm c}$ is defined as
\begin{equation}\label{eq:T_c}
    T_{\rm c}(R, U) = \left( \frac{{\lambda_{\rm air} f_{\rm h}(U) } }{4R \sigma}\right)^{1/3},
\end{equation}
with $\sigma$ the Stefan-Boltzmann constant and $f_{\rm h}(U)$ an increasing function of the air velocity $U$.
As explained before, imposing an air flow leads to $T_{\rm dry}\ll T_{\rm c}$ for typically a few meters per second with $R$ of the order of the centimeter.
In this case, the measurement becomes insensitive to the air flow.

Often, the connection between the evaporation of spherical drops and the psychrometer is not established, although the principle is the same.
In contrast to small water droplets of diameters of typically 100 $\mu$m and below, the theory of the psychrometer is more complete in the sense that the interplay between radiation and air flow is crucial for a perfect physical understanding of the instrument.

\subsection{Foam film}
The temperature of foam films can be measured by using two thermocouples \cite{Boulogne2022}.
The first one is deformed to make a circular frame, while the second one is unmodified to measure the room temperature.
Both are connected to an electronic measuring device and a a foam film is produced by dipping the frame into a solution and then pulling it out again.
The recorded temperature indicates a rapid decrease in time of the temperature, reaching a minimum within a few tens of seconds, and then slowly increasing back to room temperature.
Indeed, the thermocouple measures the temperature of the Plateau border of the foam film, which is expected to be identical to the foam film temperature.
The initial decrease is simply due to the transition between the room temperature at which the solution in the vial is, and the temperature of the foam film, which plays the role of the wet bulb  in this experiment.
We demonstrated that the minimum of temperature can be explained quantitatively by a model in the spirit of the equations derived in Section~\ref{sec:psychro}.
The second regime is attributed to glycerol, whose concentration increases as the evaporation proceeds, decreasing the evaporation rate, and thus the cooling effect.

For foam films of 12 mm in diameter, the temperature difference can reach 8~$^\circ$C below the room temperature.
So far,  the temperature distribution remains unknown, which raises the question of the existence of temperature gradients in the film.
Although apparently similar to the experiment in the thin film balance presented in the Section \ref{ssec:pure_film}, the thermal boundary conditions are different.
In the thin film balance, the outer edge, by its size and thermal capacity, is a thermostat for the film.
For foam films on a frame, evaporation also occurs at the edge lowering the temperature, as measured by the thermocouple.
Therefore, the conclusions on the Marangoni flows obtained with the thin film balance experiments must be reconsidered in the geometry on a frame.

\begin{figure}
    \centering

    \includegraphics[width=.8\linewidth]{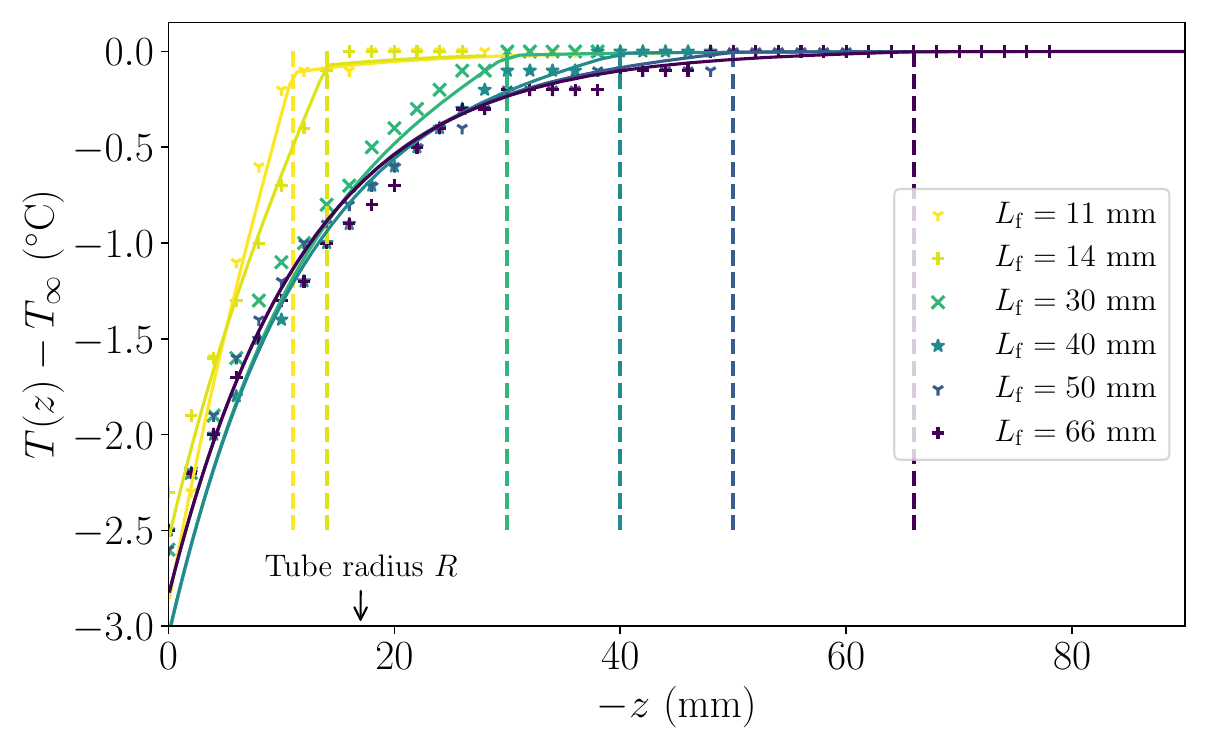}
    \includegraphics[width=.18\linewidth]{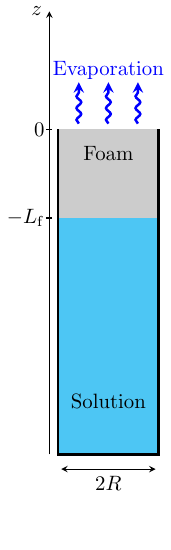}
    \caption{
    Temperature profiles $T(z) - T_\infty$ in the center of a foam column of thickness $L_{\rm f}$ evaporating from the top as represented in the schematic.
    The dry bulb temperature corresponds to $T_\infty$ and $T(z)$ is the temperature measured in the column at different depths.
    The solid lines correspond to an analytical model.
The graph is extracted from Ref.~\cite{Boulogne2023}.}
    \label{fig:temperature}
\end{figure}

\subsection{Foams}
The foam column is one of the standard geometries to quantify the foam stability, known as the Ross-Miles test \cite{Ross1941}.
By bubbling air through a needle in a cylinder partially filled with a mixture of water and dish-washing soap, a foam column of monodisperse bubbles reaching the rim is let to evaporate.
We expect a cooling at this interface and thus, a temperature gradient along the foam column \cite{Boulogne2023}.
The temperature profile at the center of the column is measured with a thermocouple as shown in Figure \ref{fig:temperature} for different foam thicknesses.
Due to the thermal insulation properties of foams, the temperature gradient spans over a macroscopic length scale.
For a foam column of radius $R$ and length $L_{\rm f}$, a 1D model predicts a temperature profile in the foam that reads

\begin{equation}
T_{\rm f}(z) =  T_\infty -\Delta T^\star \left[  \alpha_{\rm f} e^{-\sqrt{3} z/R} - \beta_{\rm f}  e^{\sqrt{2} z/R} \right]
\end{equation}
where $\alpha_{\rm f}$ and $\beta_{\rm f}$ are two coefficients, which are dependent on the aspect ratio $L_{\rm f}/R$ and the ratio of liquid and foam thermal conductivities.
A similar equation can be obtained in the liquid that we omit here for the sake of simplicity.

The temperature difference $\Delta T^\star$ is the difference between the room temperature and the temperature of the interface.
As presented in Section \ref{sec:psychro}, $\Delta T^\star$ is obtained by a psychrometric equation (Eq.~\ref{eq:psychrometric_equation}), for which the psychrometer coefficient is calculated by modeling this specific geometry \cite{Boulogne2023}.
As a result, for slender columns, the characteristic length scale of the temperature gradient is about the column radius.
This prediction is confirmed by experimental measurements of the temperature profile in the center of the tube with a thermocouple as shown in figure \ref{fig:temperature}.

Therefore, for both foam films and foams, significant temperature variations induced by evaporation are measured.
So far, the connection between the cooling and the stability is not established yet and remains to be done.

%
%
\section{Perspectives and open questions}\label{sec:perspectives}

\subsection{Summary of the state-of-the-art}

In this article, we presented key results to understand the role of evaporation on foam film and foam stability.
To do so, we proposed a short and non-comprehensive review on evaporation, limiting ourselves to the case of drops as we think that this literature is a precious helper for the current questions in the foam community.
We started by a presentation of the well-established results from a historical perspective on the evaporation of spherical drops for which the heat exchange and the air flow are considered to derive the droplet lifetime.
Then, we showed that the sessile drop is another thoroughly studied geometry in the context of the coffee-stain effect.
Depending on the involved materials, heat exchange with the substrate can create temperature gradients responsible for Marangoni flows.
So far, the number of studies devoted to these thermocapillary flows is limited and the agreement of the models with the experimental measurement is mainly qualitative.
Marangoni flows can also be triggered by gradients of liquid composition at the interface.
Naturally, evaporating liquids containing surfactant molecules are prone to generate eddies but also binary mixtures.
A rich variety of behaviors is obtained with binary mixtures since the volatility, the variation of liquid density and surface tension with the liquid composition are at play.

In Section~\ref{sec:thinning}, we presented a reading of the scientific works on foam film and foam stability to demonstrate that evaporation is one of the key aspects.
Via a thermocapillary flow, it is well-established that evaporation helps to stabilize pure liquid films produced in a thin film balance provided that the liquid flow is not impeded by viscous resistance.
Nevertheless, the temperature field has not been directly measured and this will be required for more precise descriptions of the experiments.
Interestingly, the combinations of thermocapillary and solutocapillary flows can either strengthen or decrease the stability, which depends on the nature of the solutes.
This observation demonstrates a complex interplay of Marangoni flows where the geometry is relevant in evaporation and heat transfer, and where the chemical composition is crucial as well for the surface tension gradients.
On foam films stabilized with non-volatile surfactants, experimental results suggest that evaporation decreases foam film and foam lifetimes.
This is mainly explained by a film thinning induced by evaporation in foam films.
So far, the role of temperature gradients seems to be overlooked for no specific reason.
In contrast, volatile surface active molecules are found to stabilize foam films  either for vertical films or surface bubbles.
Early studies attributed the effect to temperature gradients while more recent ones are rather considering variations of solute concentrations.
Future works must consider both mechanisms as they are susceptible to contribute to the liquid flow.

For the aforementioned reasons, we started to quantify and model the temperature of soapy objects in some particular geometries, as presented in Section~\ref{sec:temperature}.
To do so, we revisited the working principle of the psychrometer, a somewhat forgotten instrument designed to measure the relative humidity based on the temperature of a wet bulb thermometer.
To model the wet bulb, the effect of heat transfer by radiation must be taken into account, which is negligible in the case of small drops that we reviewed previously, but which is relevant in many situations on foams due to their characteristic size.

Measurements on foam films and foams indicate that temperature variations in space and time are measurable.
As evaporation depends on the geometry, careful analysis must be performed for each problem.
Nevertheless, we can affirm that thermal effects must be considered as a potential key parameter in the foam film stability.

\subsection{Perspectives}

Beyond the questions pointed out throughout the review of the literature, we identified the following perspectives that will contribute to make progress in the coming years.

\paragraph{The need for numerical modeling}
Marangoni flows play an important role in the temperature field in evaporating drops.
This question must be also addressed in foam films and foams, which will also depend on the system.
Volatile, soluble and insoluble surface active molecules are likely exhibiting drastic different behaviors and must be considered separately.
The challenge is to combine in a single model the evaporation,   triggering thermal and solutal Marangoni flows retro-acting on evaporation.
Except for simple geometries and particular cases, the complexity of these problems prevent predictions with simple analysis and requires the use of numerical computations  to unveil these intricate mechanisms.
To date, a comprehensive study combining thermal and solutal Marangoni flows in evaporating drops and foam films remains to be done.

\paragraph{Tailored experimental designs for simplified system modeling}

 The case of insoluble surfactant with respect to evaporation is particularly interesting.
Rideal observed in 1925 that a monolayer of fatty acids covering the interface of water reduces evaporation \cite{Rideal1925}.
The evaporation resistance follows an Arrhenius-type law \cite{Langmuir1943}, increasing with the chain length of interfacial molecules \cite{Barnes1986,Mer1962}, and with the surface pressure.
Insoluble surfactants can be an interesting class of systems for tuning the evaporation rate.
Additionally, the insolubility facilitates the theoretical developments regarding the equations on the transport of surfactants and Marangoni flows \cite{Karapetsas2016}.
In terms of applications, the use of these insoluble surfactants was considered as a strategy for limiting the evaporation of large scale outdoor water reservoirs \cite{Mer1962}.
However, the wind and impurities have been found to affect deleteriously the evaporation resistance \cite{Mer1965}.
The underlying physical mechanisms remain to be elucidated.

Besides insoluble surfactants, an approach where temperature gradients are imposed artificially instead of the evaporation can be useful to simplify the description of the problem \cite{Poulain2018a,Nath2022}.

\paragraph{Additional difficulties on foam films compared to drops}

Remarkably, some observations on vertical films attributed convection in liquid film to temperature gradients caused by evaporation \cite{Skotheim2000}, but this was not developed further.
Additionally, air convection can be either natural or forced, which can be particularly important in some applications.
Natural convection occurs when the Grashof number defined as the ratio of the buoyancy forces and the viscous forces, is large.
Under ambient conditions, for water,  the deviation from a purely diffusive evaporative flux becomes significant for a length scale of the evaporating surface above several centimeters \cite{Dollet2017,Boulogne2018a}.
Some foam films and foams are easily beyond this length scale, which means that the natural convection must be taken into account.
As explained in Section~\ref{sec:psychro}, the air convection not only increases the evaporation rate, but also the heat transfer by conduction.
In systems where natural convection is expected, imposing a forced convection can be a more convenient approach that has the advantage to screen the contribution of radiation.
On the other hand, the air flow imposes a shear stress to the liquid interface whose consequences will have to be determined \cite{Couder1989}.

Furthermore, as explained in Section~\ref{sec:temperature}, heat transfer by radiation cannot be neglected for objects beyond the centimeter length scale.
This suggests controlling the environmental conditions, not only regarding the temperature and the relative humidity, but also in terms of infrared emissions.

\paragraph{Experimental challenges}

Measuring the temperature of foam films presents significant experimental challenges due to their unique physical properties.
While thermocouples offer an initial approach, their limitations necessitate more advanced techniques such as infrared thermography.
However, thermography demands a high level of expertise to ensure accurate and reliable results.

The thinness of foam films renders them semi-transparent to infrared radiation, meaning an infrared camera captures a complex flux composed of three components: emission from the film itself, transmission from the background, and reflection from the surrounding environment.
Consequently, converting this radiative flux into an accurate temperature reading requires a detailed model of the infrared radiation interactions within the scene.
This model must account for the film's thickness, which can vary dynamically, necessitating complementary measurements—for example, using a hyperspectral camera to provide additional data.

The primary challenge lies in the integration and synchronization of multiple experimental techniques within a controlled environment.
Successfully integrating thermography within a climate-controlled chamber, alongside complementary measurement techniques, represents a critical step toward achieving precise and reliable temperature measurements in foam films.

\section{Conclusion}

From our analysis, we revealed that new connections must be established between two distinct scientific communities.
The community working on foam films and foams and more specifically on their stability has to learn from the studies on droplets about thermodynamic processes of evaporation associated with the complex transports involved in complex fluids such as binary liquids.
On the other hand, we have also seen that questions on the evaporation of droplets with Marangoni flows are still unsolved, which can benefit from the most advanced research on foam films.
We hope that this current opinion will contribute to foster new interactions within the Soft Matter physics community.

\section*{Declaration of interest}
Declarations of interest: none.


\bibliography{biblio}

\begin{thebibliography}{10}

\bibitem{Bikerman1973}
J.J. Bikerman.
\newblock {\em Foams}.
\newblock Springer, 1973.

\bibitem{Stevenson2012}
P.~Stevenson.
\newblock {\em Foam Engineering: Fundamentals and Applications}.
\newblock 2012.

\bibitem{Pugh2016}
R.~J. Pugh.
\newblock {\em Bubble and foam chemistry}.
\newblock Cambridge University Press, 2016.

\bibitem{Wang2016}
J.~Wang, A.~V. Nguyen, and S.~Farrokhpay.
\newblock A critical review of the growth, drainage and collapse of foams.
\newblock {\em Advances in Colloid and Interface Science}, 228:55--70, 2016.

\bibitem{Langevin2019}
D.~Langevin.
\newblock Coalescence in foams and emulsions: Similarities and differences.
\newblock {\em Current Opinion in Colloid \& Interface Science}, 44:23--31,
  2019.

\bibitem{SaintJalmes2006}
A.~Saint-Jalmes.
\newblock Physical chemistry in foam drainage and coarsening.
\newblock {\em Soft Matter}, 2:836--849, 2006.

\bibitem{Tobin2011}
S.~T. Tobin, A.~J. Meagher, B.~Bulfin, M.~Möbius, and S.~Hutzler.
\newblock {A public study of the lifetime distribution of soap films}.
\newblock {\em Amer. J. Phys.}, 79(8):819--824, 08 2011.

\bibitem{Pasquet2022a}
M.~Pasquet, L.~Wallon, P.-Y. Fusier, F.~Restagno, and E.~Rio.
\newblock An optimized recipe for making giant bubbles.
\newblock {\em Eur. Phys. J. E}, 45(12):101, 2022.

\bibitem{Li2012a}
X.~Li, S.~I. Karakashev, G.~M. Evans, and P.~Stevenson.
\newblock Effect of environmental humidity on static foam stability.
\newblock {\em Langmuir}, 28(9):4060--4068, 2012.

\bibitem{Hu2006}
H.~Hu and R.~G. Larson.
\newblock Marangoni {Effect} {Reverses} {Coffee}-{Ring} {Depositions}.
\newblock {\em The Journal of Physical Chemistry B}, 110(14):7090--7094, April
  2006.

\bibitem{Dehaeck2014}
S.~Dehaeck, A.~Rednikov, and P.~Colinet.
\newblock Vapor-based interferometric measurement of local evaporation rate and
  interfacial temperature of evaporating droplets.
\newblock {\em Langmuir}, 30(8):2002--2008, 2014.

\bibitem{Langmuir1918}
I.~Langmuir.
\newblock The evaporation of small spheres.
\newblock {\em Phys. Rev.}, 12(5):368--370, 1918.

\bibitem{Fuchs1947}
N.~Fuchs.
\newblock Concerning the velocity of evaporation of small droplets in a gas
  atmosphere.
\newblock {\em Physikalische Zeitschrift der Sowjetunion}, 6(NACA-TM-1160),
  1947.

\bibitem{Ranz1952}
W.~E. Ranz and W.R. Marshall.
\newblock Evaporation from drops: Part {I}.
\newblock {\em Chem. Eng. Prog.}, 48(3):141--146, 1952.

\bibitem{Ranz1952a}
W.~E. Ranz and W.R. Marshall.
\newblock Evaporation from drops: Part {II}.
\newblock {\em Chem. Eng. Prog.}, 48(4):173--181, 1952.

\bibitem{Frossling1938}
N.~Fr\"ossling.
\newblock Uber die verdunstung fallernder tropfen.
\newblock {\em Gerlands Beitr. Geophys.}, 52:170--216, 1938.

\bibitem{Andreas1989}
E.~L. Andreas.
\newblock Thermal and size evolution of sea spray droplets.
\newblock Technical report, 1989.

\bibitem{Andreas1995}
E.~L. Andreas.
\newblock The temperature of evaporating sea spray droplets.
\newblock {\em J. Atmos. Sci.}, 52(7):852--862, April 1995.

\bibitem{Sobac2015}
B.~Sobac, P.~Talbot, B.~Haut, A.~Rednikov, and P.~Colinet.
\newblock A comprehensive analysis of the evaporation of a liquid spherical
  drop.
\newblock {\em J. Colloid Interface Sci.}, 438:306 -- 317, 2015.

\bibitem{Netz2020}
R.~R. Netz.
\newblock Mechanisms of airborne infection via evaporating and sedimenting
  droplets produced by speaking.
\newblock {\em J. Phys. Chem. B}, 124(33):7093--7101, 2020.

\bibitem{Netz2020a}
R.~R. Netz and W.~A. Eaton.
\newblock Physics of virus transmission by speaking droplets.
\newblock {\em Proc. Natl. Acad. Sci.}, 117(41):25209--25211, 2020.

\bibitem{Corpart2023a}
M.~Corpart, F.~Restagno, and F.~Boulogne.
\newblock Analytical prediction of the temperature and the lifetime of an
  evaporating spherical droplet.
\newblock {\em Colloids and Surfaces A}, 2023.

\bibitem{Deegan1997}
R.~D. Deegan, O.~Bakajin, T.~F. Dupont, G.~Huber, S.~R. Nagel, and T.~A.
  Witten.
\newblock Capillary flow as the cause of ring stains from dried liquid drops.
\newblock {\em Nature}, 389(6653):827--829, October 1997.

\bibitem{Boulogne2017a}
F.~Boulogne, F.~Ingremeau, and H.~A. Stone.
\newblock Coffee-stain growth dynamics on dry and wet surfaces.
\newblock {\em J. Phys.: Condens. Matter}, 29(7):074001, 2017.

\bibitem{Dambrosio2023}
H.-M. D’Ambrosio, S.~K. Wilson, A.~W. Wray, and B.~R. Duffy.
\newblock The effect of the spatial variation of the evaporative flux on the
  deposition from a thin sessile droplet.
\newblock {\em Journal of Fluid Mechanics}, 970:A1, 2023.

\bibitem{Deegan2000}
R.~D. Deegan.
\newblock Pattern formation in drying drops.
\newblock {\em Phys. Rev. E}, 61:475--485, Jan 2000.

\bibitem{Ristenpart2007}
W.~D. Ristenpart, P.~G. Kim, C.~Domingues, J.~Wan, and H.~A. Stone.
\newblock Influence of substrate conductivity on circulation reversal in
  evaporating drops.
\newblock {\em Phys. Rev. Lett.}, 99:234502, Dec 2007.

\bibitem{Gelderblom2022}
H.~Gelderblom, C.~Diddens, and A~Marin.
\newblock Evaporation-driven liquid flow in sessile droplets.
\newblock {\em Soft Matter}, 18:8535--8553, 2022.

\bibitem{Marin2016}
A.~Marin, R.~Liepelt, M.~Rossi, and C.~J. Kähler.
\newblock Surfactant-driven flow transitions in evaporating droplets.
\newblock {\em Soft Matter}, 12:1593--1600, 2016.

\bibitem{Still2012}
T.~Still, P.~J. Yunker, and A.~G. Yodh.
\newblock Surfactant-induced marangoni eddies alter the coffee-rings of
  evaporating colloidal drops.
\newblock {\em Langmuir}, 28(11):4984--4988, 2012.

\bibitem{Guena2007a}
G.~Guéna, C.~Poulard, and A.M. Cazabat.
\newblock Evaporating drops of alkane mixtures.
\newblock {\em Colloids and Surfaces A: Physicochemical and Engineering
  Aspects}, 298(1):2--11, 2007.

\bibitem{Fournier1992}
J.~B. Fournier and A.~M. Cazabat.
\newblock Tears of wine.
\newblock {\em EPL}, 20(6):517, 1992.

\bibitem{Diddens2017}
C.~Diddens, H.~Tan, P.~Lv, M.~Versluis, J. G. M. Kuerten, X.~Zhang, and
  D.~Lohse.
\newblock Evaporating pure, binary and ternary droplets: thermal effects and
  axial symmetry breaking.
\newblock {\em Journal of Fluid Mechanics}, 823:470–497, 2017.

\bibitem{Kim2016}
H.~Kim, F.~Boulogne, E.~Um, I.~Jacobi, E.~Button, and H.~A. Stone.
\newblock Controlled uniform coating from the interplay of {M}arangoni flows
  and surface-adsorbed macromolecules.
\newblock {\em Phys. Rev. Lett.}, 116:124501, 2016.

\bibitem{Diddens2021}
C.~Diddens, Y.~Li, and D.~Lohse.
\newblock Competing {M}arangoni and {R}ayleigh convection in evaporating binary
  droplets.
\newblock {\em Journal of Fluid Mechanics}, 914:A23, 2021.

\bibitem{Plateau1873}
J.~A.~F. Plateau.
\newblock {\em Statique exp{\'e}rimentale et th{\'e}orique des liquides soumis
  aux seules forces mol{\'e}culaires}.
\newblock Gauthier-Villars, 1873.

\bibitem{Novev2017}
J.~K. Novev, N.~Panchev, and R.~I. Slavchov.
\newblock Evaporating foam films of pure liquid stabilized via the thermal
  marangoni effect.
\newblock {\em Chem. Eng. Sci.}, 171:520--533, 2017.

\bibitem{Sheludko1967}
A.~Sheludko.
\newblock Thin liquid films.
\newblock {\em Advances in Colloid and Interface Science}, 1(4):391--464, 1967.

\bibitem{Yaminsky2006}
V.V. Yaminsky.
\newblock Bubble vortex at surfaces of evaporating liquids.
\newblock {\em Journal of Colloid and Interface Science}, 297(1):251--260,
  2006.

\bibitem{Yaminsky2010}
V.~V. Yaminsky, S.~Ohnishi, E.~A. Vogler, and R.~G. Horn.
\newblock Stability of aqueous films between bubbles. {P}art 1. {T}he effect of
  speed on bubble coalescence in purified water and simple electrolyte
  solutions.
\newblock {\em Langmuir}, 26(11):8061--8074, June 2010.

\bibitem{Yaminsky2010a}
V.~V. Yaminsky, S.~Ohnishi, E.~A. Vogler, and R.~G. Horn.
\newblock Stability of aqueous films between bubbles. {P}art 2. {E}ffects of
  trace impurities and evaporation.
\newblock {\em Langmuir}, 26(11):8075--8080, 2010.

\bibitem{Menesses2019}
M.~Menesses, M.~Roch\'e, L.~Royon, and J.~C. Bird.
\newblock Surfactant-free persistence of surface bubbles in a volatile liquid.
\newblock {\em Phys. Rev. Fluids}, 4:100506, Oct 2019.

\bibitem{Poulain2018}
S.~Poulain, E.~Villermaux, and L.~Bourouiba.
\newblock Ageing and burst of surface bubbles.
\newblock {\em J. Fluid Mech.}, 851:636–671, 2018.

\bibitem{Miguet2021}
J.~Miguet, F.~Rouyer, and E.~Rio.
\newblock The life of a surface bubble.
\newblock {\em Molecules}, 26(5), 2021.

\bibitem{Miguet2019}
J.~Miguet, M.~Pasquet, F.~Rouyer, Y.~Fang, and E.~Rio.
\newblock Stability of big surface bubbles: Impact of evaporation and bubbles
  size.
\newblock {\em Soft Matter}, 16:1082--1090, 2019.

\bibitem{Champougny2018}
L.~Champougny, J.~Miguet, R.~Henaff, F.~Restagno, F.~Boulogne, and E.~Rio.
\newblock Influence of evaporation on soap film rupture.
\newblock {\em Langmuir}, 34(10):3221--3227, 2018.

\bibitem{Pasquet2022}
M.~Pasquet, F.~Boulogne, J.~Saint-Anna, F.~Restagno, and E.~Rio.
\newblock Impact of physical-chemistry on the film thinning in surface bubbles.
\newblock {\em Soft Matter}, 18:4536--4542, 2022.

\bibitem{Roux2022}
A.~Roux, A.~Duchesne, and M.~Baudoin.
\newblock Everlasting bubbles and liquid films resisting drainage, evaporation,
  and nuclei-induced bursting.
\newblock {\em Phys. Rev. Fluids}, 7:L011601, Jan 2022.

\bibitem{Li2010}
X.~Li, R.~Shaw, and P.~Stevenson.
\newblock Effect of humidity on dynamic foam stability.
\newblock {\em Int. J. Miner. Process.}, 94(1–2):14 -- 19, 2010.

\bibitem{Talmud1931}
D.~Talmud and S.~Suchowolskaya.
\newblock Stabilitât des elementaren schaurnes.
\newblock {\em Z. physik. Chem.}, 154{A}:277, 1931.

\bibitem{Jones1967}
M.~N. Jones and K.~J. Mysels.
\newblock Juvenescent soap films. the evaporative stabilization of a film by a
  volatile surfactant.
\newblock {\em J. Am. Oil Chem. Soc.}, 44(5):284--288, 1967.

\bibitem{Pigeonneau2012}
F.~Pigeonneau, H.~Kočárková, and F.~Rouyer.
\newblock Stability of vertical films of molten glass due to evaporation.
\newblock {\em Colloids and Surfaces A: Physicochemical and Engineering
  Aspects}, 408:8--16, 2012.

\bibitem{Lorenceau2020}
E.~Lorenceau and F.~Rouyer.
\newblock Lifetime of a single bubble on the surface of a water and ethanol
  bath.
\newblock {\em Phys. Rev. Fluids}, 5(6):063603, 2020.

\bibitem{Nath2022}
S.~Nath, G.~Ricard, P.~Jin, A.~Bouillant, and D.~Quéré.
\newblock Thermal marangoni bubbles.
\newblock {\em Soft Matter}, 18:7422--7426, 2022.

\bibitem{Corpart2023b}
M.~Corpart, F.~Restagno, and F.~Boulogne.
\newblock Measuring relative humidity from evaporation with a wet-bulb
  thermometer: {T}he psychrometer.
\newblock {\em American Journal of Physics}, 91(11), 2023.

\bibitem{Playfair1997}
J.~Playfair.
\newblock Biographical account of the late {Dr James Hutton, F. R. S. Edin}.
\newblock {\em Earth and Environmental Science Transactions of the Royal
  Society of Edinburgh}, 88(S1):39–99, 1997.

\bibitem{Ivory1822}
J.~Ivory.
\newblock {XVI.} {O}n the hygrometer by evaporation.
\newblock {\em The Philosophical Magazine}, 60(292):81--88, 1822.

\bibitem{Boulogne2022}
F.~Boulogne, F.~Restagno, and E.~Rio.
\newblock Measurement of the temperature decrease in evaporating soap films.
\newblock {\em Phys. Rev. Lett.}, 129:268001, 2022.

\bibitem{Boulogne2023}
F.~Boulogne, E.~Rio, and F.~Restagno.
\newblock Evaporation-induced temperature gradient in a foam column.
\newblock {\em Langmuir}, 39(40):14256--14262, 2023.

\bibitem{Ross1941}
J.~Ross and G.~D. Miles.
\newblock An apparatus for comparison of foaming properties of soaps and
  detergents.
\newblock {\em Oil \& Soap}, 18(5):99--102, 1941.

\bibitem{Rideal1925}
E.~K. Rideal.
\newblock On the influence of thin surface films on the evaporation of water.
\newblock {\em J. Phys. Chem.}, 29(12):1585--1588, December 1925.

\bibitem{Langmuir1943}
I.~Langmuir and V.~J. Schaefer.
\newblock Rates of evaporation of water through compressed monolayers on water.
\newblock {\em J. Franklin Inst. B}, 235(2):119 -- 162, 1943.

\bibitem{Barnes1986}
G.T. Barnes.
\newblock The effects of monolayers on the evaporation of liquids.
\newblock {\em Adv. Colloid Interface Sci.}, 25:89--200, 1986.

\bibitem{Mer1962}
V.K La~Mer, editor.
\newblock {\em Retardation of Evaporation by Monolayers: Transport Processes}.
\newblock Academic Press, 1962.

\bibitem{Karapetsas2016}
G.~Karapetsas, K.~C. Sahu, and O.~K. Matar.
\newblock Evaporation of sessile droplets laden with particles and insoluble
  surfactants.
\newblock {\em Langmuir}, 32(27):6871--6881, July 2016.

\bibitem{Mer1965}
V.~K.~La Mer and T.~W. Healy.
\newblock Evaporation of water: Its retardation by monolayers.
\newblock {\em Science}, 148(3666):36--42, 1965.

\bibitem{Poulain2018a}
S.~Poulain and L.~Bourouiba.
\newblock Biosurfactants change the thinning of contaminated bubbles at
  bacteria-laden water interfaces.
\newblock {\em Phys. Rev. Lett.}, 121:204502, Nov 2018.

\bibitem{Skotheim2000}
J.~M. Skotheim and J.~W. Bush.
\newblock Evaporatively driven convection in a draining soap film.
\newblock {\em Physics of Fluids}, 12(9):S3--S3, 2000.

\bibitem{Dollet2017}
B.~Dollet and F.~Boulogne.
\newblock Natural convection above circular disks of evaporating liquids.
\newblock {\em Phys. Rev. Fluids}, 2:053501, 2017.

\bibitem{Boulogne2018a}
F.~Boulogne and B.~Dollet.
\newblock Convective evaporation of vertical films.
\newblock {\em Soft Matter}, 14:1665--1671, 2018.

\bibitem{Couder1989}
Y.~Couder, J.M. Chomaz, and M.~Rabaud.
\newblock On the hydrodynamics of soap films.
\newblock {\em Physica D: Nonlinear Phenomena}, 37(1):384--405, 1989.

\end{thebibliography}

\bibliographystyle{unsrt}



\end{document}